\documentclass[aps,notitlepage,a4paper,superscriptaddress,11pt]{article}

\usepackage[utf8]{inputenc}
\usepackage[dvips]{color}
\usepackage[toc]{appendix}
\usepackage[hang,flushmargin]{footmisc} 
\usepackage{titlesec,titletoc,ntheorem-hyper,authblk,abstract,latexsym,microtype,booktabs,amsmath,cleveref,fancyhdr,wrapfig,array,amsfonts, amssymb,geometry,appendix,cite}
\usepackage{dsfont}

\newcommand{\be}{\begin{equation}}
\newcommand{\ee}{\end{equation}}
\newcommand{\bea}{\begin{eqnarray}}
\newcommand{\eea}{\end{eqnarray}}
\newcommand\blfootnote[1]{%
  \begingroup
  \renewcommand\thefootnote{}\footnote{#1}%
  \addtocounter{footnote}{-1}%
  \endgroup
}

\usepackage{graphicx,lipsum}
\usepackage{caption}
\usepackage{subfigure}
\usepackage{amsmath}
\usepackage{amsfonts}
\usepackage{amssymb}
\usepackage{stackengine}
\usepackage{mathtools}
\numberwithin{equation}{section}
\usepackage{titlesec}
\usepackage{sectsty}
\sectionfont{\centering}

\numberwithin{subcase}{case}
\usepackage{framed}
\allowdisplaybreaks
\usepackage[nottoc]{tocbibind}
\usepackage[free-standing-units]{siunitx}
\usepackage{helvet}
\usepackage{url}
\usepackage{titletoc}
\usepackage{blindtext}
\usepackage[gen]{eurosym}

\title{Position-dependent mass Dirac equation and local Fermi velocity}
\author{Rahul Ghosh}
\affil{Physics Department, Shiv Nadar University, Gautam Buddha Nagar, \\
Uttar Pradesh 203207, India}

\begin{document}

\maketitle

\begin{abstract}
Abstract: We present a new approach to study (1+1)-dimensional Dirac equation in the background of an effective mass $M$ by exploiting the possibility of a position-dependent Fermi velocity $v_F$. We explore the resulting structure of the coupled equations and arrive at an interesting constraint of $M$ turning out to be the inverse square of $v_F$. We address several solutions of the effective potential that include the free particle, shifted harmonic oscillator, Morse potential and CPRS potential.
\end{abstract}

\blfootnote{E-mails: rg928@snu.edu.in}

\section{Introduction}

Dirac equation is recognized as one of the finest achievements in twentieth-century physics. It is a relativistic equation in its own right that describes the dynamics of spin one-half particles. It has been investigated from various perspectives of which the list is long \cite{tha}. As a sample survey, let us mention that its study was made on a hyperbolic graphene surface under perpendicular magnetic fields \cite{neg}, in connection with confinement in two dimensions of nonuniform magnetic fields for massless fermions \cite{dow}, and from the point of view of first-order intertwining operators \cite{fer}. Furthermore, Dirac equation has also received attention following the production of graphene crystals as two-dimensional, single carbon
atom sheets (see, for instance, \cite{nov}) and reports were presented for the understanding of the electronic properties of the charge carriers by two-dimensional zero-mass Dirac particles \cite{cast, gall1}. It is also known that position-dependent Fermi velocity could be induced from a nonuniform strain in graphene as scanning tunnelling spectroscopy experiments reveal \cite{juan, yan, jang}. We might remark that the electron concentration in the conduction band is greatly affected by the Fermi energy which is much smaller than that of a metal The idea of position-dependent Fermi velocity also gained acceptance when a gap formation was noticed in graphene physics \cite{gui}. A number of theoretical investigations incorporating spatially varying Fermi velocity have also been carried out \cite{panella2012, mus2013, oliva} including a recent one studying electronic transport in two-dimensional strained Dirac materials \cite{phan}. We note in passing that 

Unconventional aspects of supersymmetry were shown to exist in graphene and for its associated problems \cite{gall2}. 
Previously, interpretation in the realm of supersymmetric quantum mechanics \cite{rau, jun1} has been given.  More recently, the most general form of the one-dimensional Dirac Hamiltonian in the presence of scalar and pseudoscalar potentials was written down and analysed \cite{bag2021} (and references therein) to look for the role of the intertwining relations. 

In the underlying structure of Dirac equation, the present work is devoted to extending some of its aspects in the context of  position-dependent mass (PDM) and fermi velocity. 
We have concentrated on the study of the coupled equations, resulting from the spinor interpretation of the 2-component wavefunctions, by considering the effects of space
dependent mass as well as Fermi velocity. The aim of this paper is to study (1+1)-dimensional Dirac equation in a PDM theory where Fermi velocity depends on the position.

The paper is organised as follows.   
In section 2 we introduce the idea of localised Fermi velocity(LFV). In section 3 we explore the PDM issue by comparing with the von Roos form of the effective Hamiltonian. We also explore here the general behaviour of the effective potential. In section 4 we address several classes of examples following from our scheme.  Finally a summary is presented in section 5. 

\section{(1+1)-Dirac equation and the PDM}

The $(1+1)$-Dirac Hamiltonian \cite{alh2011,jun2020} can be expressed in several equivalent forms. One such that suits our convenience is given by 
\begin{equation}\label{H_D1}
  H_D = v_f \sigma_x p_x  +\sigma_y W(x)  +\sigma_z  mv_f^2  +\mathds{1} V(x) 
\end{equation}
where $\mathds{1}$ is the block-diagonal unit matrix and the associated potential, in general, consists of the electrostatic contribution $V(x)$, along with $W(x)$ which corresponds to the pseuodoscalar contribution. We shall see that the latter acts as the superpotential of the system. The standard expressions of the Pauli matrices are known to be
\begin{equation}
\sigma_x = \left( \begin{array}{ccc|c} 0 & 1  \\ 1 & 0  \end{array} \right), \quad \sigma_y = \left( \begin{array}{ccc|c} 0 & -i  \\ i & 0  \end{array} \right), \quad \sigma_z = \left( \begin{array}{ccc|c} 1 & 0  \\ 0 & -1  \end{array} \right) 
\end{equation}
 The solution of the non-relativistic version of the Dirac equation following from $ H_D $ is known for long in the context of an external magnetic field \cite{rab} and for massless particles in (2+1)-dimensions \cite{chi} by operating on a two-component spinor wave function. In the framework of first-order intertwining relations the role of the electrostatic potential $V(x)$ is ineffective \cite{ish}.
 
 In the context of treating Dirac equation in the background of PDM,  we adopt the von Roos method of converting the kinetic energy operator to a Hermitian form by introducing a set of three ambiguity parameters. In the theoretical studies a number of cases of different choices of ambiguity parameters have been investigated. These correspond to the schemes of Ben Daniel and Duke \cite{ben} $(\alpha=\gamma=0, \beta =-1)$, Bastard \cite{bas} $(\alpha=-1,\gamma=\beta=0)$, Zhu and Kroemar \cite{zhu} $(\alpha=\gamma=-\frac{1}{2})$, and redistributed model \cite{li} $(\beta=\gamma=-\frac{1}{2}, \alpha=0)$. 

We will focus\footnote{Of course, we can readily extend to other PDM types by changing the values of the ambiguity parameters.} on the Ben Daniel and Duke model to seek correspondence of the resulting Schr\"{o}dinger form for the two cases of the Dirac Hamiltonian when the latter is applied on the spinor wave equation. The Ben Daniel and Duke model is a physically interesting model and has found, among other places, applications in nanowire semiconductor heterostructure (\cite{will} and references therein). In fact, one can construct eigenstates and spectral values of nanowires having different cross-sectional shape and/or varying composition.  

A combination of Fermi velocity and the mass function carries the bulk information of the material properties of the Dirac particle \cite{mele}. Indeed, these give rise to the possibility of producing a heterostructure. Since the Fermi velocity can vary from material to material, such a scenario is comprehensible in condensed matter physics \cite{per2009}. In the following section we will enquire how a localized Fermi velocity influences the Dirac equation.

 \section{Localized Fermi Velocity}

Let us modify (\ref{H_D1}) by incorporating a modulated velocity in 1-dimensional heterostructure as follows 
\begin{gather} \label{H_D}
  H_D = \sqrt{v_f(x)} \sigma_x p_x \sqrt{v_f(x)} +\sigma_y W(x)  +\sigma_z  m(x) v_f^2 (x)   
\end{gather}
where following \cite{per2009} the position dependence in the mass and Fermi velocity has been taken into account. Then the following matrix structure of $H_D$ emerges

\begin{gather}\label{H_Dmatrix}
     H_D = \left( \begin{array}{cc} m v_f^2+V  &  -i\hbar \sqrt{v_f} \partial \sqrt{v_f} -iW  \\ -i\hbar \sqrt{v_f} \partial \sqrt{v_f}  + i W & - m v_f^2+V   \end{array} \right)
\end{gather}
In (\ref{H_Dmatrix}), the quantities $m$, $v_f$, $V$ and $W$ are arbitrary function of $x$. Applying $H_D$ on a two-component spinor wavefunction having components $ (\psi_+, \psi_-)^T$, the  Dirac equation assumes the form 
\begin{gather}
     \left( \begin{array}{cc} m v_f^2+V  &  -i\hbar \sqrt{v_f} \partial \sqrt{v_f} -iW  \\ -i\hbar \sqrt{v_f} \partial \sqrt{v_f}  + i W & - m v_f^2+V   \end{array} \right)  \left(
     \begin{array}{cc} \psi_+  \\ \psi_-  \end{array} \right)  = E  \left(
     \begin{array}{cc} \psi_+  \\ \psi_-  \end{array} \right)
\end{gather}
where $E$ is the energy eigenvalue. These result in a set of coupled equations

 \begin{gather}
 (-i\hbar \sqrt{v_f} \partial \sqrt{v_f} -i W) \psi_- = D_- \psi_+   \\
 (-i\hbar \sqrt{v_f} \partial \sqrt{v_f} +i W) \psi_+ = D_+ \psi_- 
\end{gather}
where we adopted natural units $\hbar = 1$ and $D_+ = (E+m v_f^2-V)$ and $D_- = (E-m v_f^2-V)$. 
Through disentanglement we get for the upper component
\begin{gather}
-\frac{v_f^2}{D_+}\frac{d^2 \psi_+}{dx^2} - \frac{d}{dx}\Big(\frac{v_f^2}{D_+}\Big) \frac{d\psi_+}{dx} + \Big[\frac{1}{D_+} \Big(W^2 -\frac{1}{4}{v'_f}^2-\frac{1}{2}v_f v''_f\Big) + v_f\frac{d}{dx} \Big(\frac{W}{D_+}\Big)     \nonumber \\ 
- \frac{1}{2}v_f v'_f \frac{d}{dx}\Big(\frac{1}{D_+}\Big)  \Big] \psi_+ = D_- \psi_+
\end{gather}
and a similar one for the lower component

\begin{gather}
-\frac{v_f^2}{D_-}\frac{d^2 \psi_-}{dx^2} - \frac{d}{dx}\Big(\frac{v_f^2}{D_-}\Big) \frac{d\psi_-}{dx} + \Big[\frac{1}{D_-} \Big(W^2 -\frac{1}{4}{v'_f}^2-\frac{1}{2}v_f v''_f\Big) -v_f\frac{d}{dx}\Big(\frac{W}{D_-}\Big)     \nonumber \\ 
- \frac{1}{2}v_f v'_f \frac{d}{dx}\Big(\frac{1}{D_-}\Big)  \Big] \psi_- = D_+ \psi_-
\end{gather}
For onward calculations we will employ Mustafa's invariant relation \cite{mus2013} $m(x)v_f^2(x)=m_0v_0^2$, where $m_0$ and $v_f = v_0$ refers to their constant values and also  set  $V(x)=0$. As a result both $D_+=E+m_0v_0^2$ and $D_-=E-m_0v_0^2$ turn out to be constant. After some straightforward rearrangement we are led to the following uncoupled equations
\begin{gather} \label{upperDirac}
- v_f^2 \frac{d^2 \psi_+}{dx^2} - \frac{d}{dx}(v_f^2) \frac{d\psi_+}{dx} + \Big[ \Big(W^2 -\frac{1}{4}{v'_f}^2-\frac{1}{2}v_f v''_f\Big) + v_f W' )\Big] \psi_+ = (E^2-m_0^2V_0^4) \psi_+     
\end{gather}
\begin{gather}\label{lowerDirac}
- v_f^2 \frac{d^2 \psi_-}{dx^2} - \frac{d}{dx}(v_f^2) \frac{d\psi_-}{dx} + \Big[ \Big(W^2 -\frac{1}{4}{v'_f}^2-\frac{1}{2}v_f v''_f\Big) - v_f W' )\Big] \psi_- = (E^2-m_0^2V_0^4) \psi_-
\end{gather}

Interestingly, if we transform the paired wavefunctions according to

\begin{gather}
    \psi_\pm(x) = \frac{1}{\sqrt{v_f(x)}} \Phi_\pm (y(x))  
\end{gather}
where the function $y(x)$ is defined by
\begin{equation}
     y(x)= \int^x \frac{dz}{v_f(z)}+ constant
\end{equation}
then the two component equations \ref{upperDirac} and \ref{lowerDirac} make over to the forms
\begin{gather}\label{newSE}
    - \frac{d^2 \Phi_\pm(y)}{dy^2}  + \Big[ \Big(W^2(x(y)) \pm v_f(x(y)) W'(x(y)) )\Big] \Phi_\pm(y) =  (E^2- m_0^2 v_0^4) \Phi_\pm(y)
\end{gather}
these can be viewed as the equations for the extended SUSY partner potentials. Indeed the reduction to their standard SUSY representations is obvious in the constant $v_f$ case.


\section{A new look at the PDM problem}

The scenario of PDM \cite{vonroos} has been widely studied in the literature. These include setting up of an extended scheme to generate the associated class of potentials \cite{bag2} and examining the consequences of a deformed shape invariance condition \cite{bag3}, a general strategy to tackle solvable potentials \cite{dut}, exploring point canonical transformation \cite{mus}, seeking new types of exact solutions for an effective mass system \cite{yes, dha}, investigating on the sphere and hyperbolic plane \cite{carinena}, looking for invariants and spectrum generating algebras \cite{ortiz}, analyzing consistency of indefinite effective mass \cite{zno}, and obtaining analytical results for PDM problems \cite{cunha}. Interest in PDM systems was triggered by the physical problems pertaining to  compositionally graded crystals \cite{Gel}, quantum dots \cite{Ser}, liquid crystals \cite{Bar} etc.

A general strategy of writing down a hermitian form of the effective-mass kinetic energy operator $\hat{T}$ for PDM guided mechanical systems was given in \cite{vonroos}.
 In terms of a position-dependent mass function $\mu (x)$, the kinetic energy operator $\hat{T}$ in this case takes the form

\begin{gather}
 \hat{T}= -\frac{\hbar^2}{4} (\mu^\eta(x) p \mu^\beta(x) p \mu^\gamma(x) + \mu^\gamma(x) p \mu^\beta(x) p \mu^\eta(x))
\end{gather}
It is found to contain a set of ambiguity parameters $\eta, \beta$ and $\gamma$ which are controlled by the relation

\begin{equation}
    \eta+\beta+\gamma = -1
\end{equation}
We emphasize that the ordering of the non-commutative momentum and mass operators have been so arranged in (3.1) that the resulting form of the kinetic energy operator $\hat{T}$ is rendered Hermitian. Although its form is not unique, other acceptable representations inevitably reduce to one of the special cases discussed earlier in the introduction and show equivalence to a reasonable accuracy \cite{Gel}.\\

On replacing $\mu(x)=\mu_0 M(x)$ the  time-independent Schr\"{o}dinger equation can be converted to   \cite{bag2}
\begin{gather} \label{pdmse}
H\phi(x)=\Big[- \frac{d}{dx}\frac{1}{M(x)}\frac{d}{dx}+V_{eff} \Big]\phi(x)=\epsilon \phi(x)
\end{gather}
where the effective potential $V_{eff}$ is seen to depend on the mass function, $M(x)$ and its derivatives
\begin{gather} \label{veffmain}
V_{eff}= \mathcal{V}(x) +\frac{1}{2}(\beta+1)\frac{M''}{M^2}-[\eta(\eta+\beta+1)+\beta+1]\frac{M'^2}{M^3}
\end{gather}
In $(3.4)$ the primes are refer to the derivatives  with respect to $x$ and $\mathcal{V}(x)$ corresponds to some given potential. Corresponding to the parameters of the Ben Daniel-Duke formalism \cite{ben}, the effective potential $V_{eff}(x)$ coincides with the system potential $\mathcal{V}(x)$.\\

It is useful to point out that it was shown by Bagchi et al \cite{bag3} several years ago that on setting $M = g^{-2} (x)$,  $(3.3)$ could be transformed to 

\begin{equation}
    H \phi (x) = \left [- \left ( \sqrt{g(x)} \frac{d}{dx} \sqrt{g(x)} \right )^2 + V_{eff} (x) \right ] \phi (x) = E \phi (x)
\end{equation}

Comparing (\ref{pdmse}) with (\ref{upperDirac}) and (\ref{lowerDirac}) we readily see that the mass function $M(x)$, for both upper (+) and lower (-) components, coincides and given by the inverse square of the Fermi velocity
\begin{gather}
 M_\pm(x)= \frac{1}{v_f^2(x)}
\end{gather}
In other words, $g (x)$ plays the role of local Fermi velocity. This is one of important realization of this work. We now proceed to show how the above result impacts the effective potential and solve for their corresponding forms.

\section{Some examples of solvable systems}

\subsection{Free Particle}
Choosing the pseudoscalar potential to be constant i.e $W(x)=\omega_0$ will correspond to the free particle case. With the following form of the LFV $v_f(x)=a^2 x^2+1$, the partner SUSY potentials assume the forms 
\begin{gather}
    V_\pm(x) = \omega_0^2
\end{gather}
These correspond to just a constant shift in the energy level 
i.e constant-potential in whole x-line. A point to notice is that if we transform $y=\frac{1}{a} \tan^{-1}(a x) \in (\frac{-\pi}{2a},\frac{\pi}{2a})$, the free-particle problem gets transformed to the case of the infinite square well potential which satisfies the Schr\"{o}dinger equation 
\begin{gather}
    - \frac{d^2 \Phi_\pm(y)}{dy^2}  + k^2 \Phi_\pm(y) =  0
\end{gather}
where $k^2=E^2-m_0^2v_0^4-\omega_0^2$. The normalized solutions are of the same forms for both the component(upper and lower) and are given by
\begin{gather}
   \Phi_n (y)= 
\begin{cases}
    \sqrt{\frac{2a}{\pi}} \sin(k_n y),     & \text{for even n } \\
    \sqrt{\frac{2a}{\pi}} \cos(k_n y),     & \text{for odd n}
\end{cases}
\end{gather}
where the wavevector $k_n= a n$ and the associated energy eigenvalues are 

\begin{equation}
E^2_n=a^2 n^2+m_0^2v_0^4+\omega_0^2
\end{equation}
For completeness we furnish the forms of the wavefunctions in the $x$-coordinate

\begin{gather}
   \psi_n (x, a)= 
\begin{cases}
    \sqrt{\frac{2a}{\pi}} \frac{\sin(n (\tan^{-1}(a x))}{\sqrt{1+a^2x^2}},     & \text{for even n } \\
    \sqrt{\frac{2a}{\pi}} \frac{\cos(n (\tan^{-1}(a x))}{\sqrt{1+a^2x^2}},     & \text{for odd n}
\end{cases}
\end{gather}
where the presence of the parameter $a$ is explicitly indicated in  the argument of the wavefunction. 

\subsection{Shifted Harmonic Oscillator}
Here we inquire in to the possibility of the pseudoscalar potential varying inversely proportional to LFV i.e $W(x)=(a e^{\alpha x}+b)$ and consider the damping form of the latter i.e. $ v_f(x)=v_0 e^{-\alpha x} $. Then the Schr\"{o}dinger equation (\ref{newSE}) is turned into the form
\begin{gather}\label{newSHO}
    - \frac{d^2 \Phi_\pm(y)}{dy^2}  + V_\pm(x(y)) \Phi_\pm(y) =  (E^2-m_0^2v_0^4) \Phi_\pm(y)     
\end{gather}
where the potential reads $ V_\pm(x) = a^2 e^{2\alpha x}+ 2abe^{\alpha x}+(b^2 \pm v_0 a \alpha)$. This is well known to be the Morse type potential. Opting for the transformation converting the full-line $(-\infty, \infty)$ to the half-line representation through $y=\frac{1}{v_0 \alpha}e^{\alpha x} \in (0, \infty)$, one readily moves over to the shifted half-harmonic oscillator $ V_\pm(y) = a^2 v_0^2 \alpha^2 y^2+ 2abv_0 \alpha y+(b^2 \pm v_0 a \alpha)$. For the concrete case $b=0$ (\ref{newSHO}) can be rewritten as
\begin{gather}
    - \frac{d^2 \Phi_\pm(y)}{dy^2}  + [\frac{1}{2}\omega^2 y^2\pm \frac{1}{\sqrt{2}}\omega] \Phi_\pm(y) =  (E^2-m_0^2v_0^4) \Phi_\pm(y)  
\end{gather}
where $\omega =\sqrt{2} v_0 a \alpha$. Both $\Phi_\pm(y)$  possess eigenfunctions described by the Hermite polynomials

\begin{gather}\label{shosol}
\Phi_n(y) = \frac{1}{\sqrt{\sqrt{\pi}2^n n!x_0}}  e^{-\frac{y^2}{2y_0^2}} H_n \left (\frac{y}{y_0} \right), \quad n = 1, 3, 5,...
\end{gather}
where $ y_0=\sqrt{\frac{2}{\omega}}$. The explicit energies for the upper and lower component are 
\begin{gather*}
    E_n^2=\omega(n+\frac{1}{2}) \pm \frac{1}{2}\omega^2 +m_0^2 v_0^4  \quad \text{where  $n=1,3,5...$}
\end{gather*}

\subsection{1-dimensional Coulomb problem}
Full-line Coulomb problem belongs to the class of singular potentials in 1-dimension. An early work on it is due to Loudon \cite{lou} who concluded that except for the ground state the remaining energies displayed a two-fold degeneracy. Later, Andrews' treatment was more or less in line with London's contention except for the interpretation of certain technical features \cite{and}. A more general treatment for $N$-dimensions was provided by Nieto \cite{nie}. A good review of the essential results can be found in \cite{pal}.

Let us consider the choice of the pseudoscalar potential as being proportional to LVF, i.e, $W(x)=(a e^{-\alpha x}-b)$. Further, we assume a damping form of the Fermi velocity, i.e., $ v_f(x)=v_0 e^{-\alpha x} $. Equation (\ref{newSE}) turns into
\begin{gather}\label{coulomb}
    - \frac{d^2 \Phi_\pm(y)}{dy^2}  + V_\pm(x(y)) \Phi_\pm(y) =  (E^2-m_0^2v_0^4) \Phi_\pm(y)     
\end{gather}
where $V_\pm(x)$ are given by 

\begin{equation}
V_\pm(x) = (a^2 \mp v_0 a \alpha) e^{-2\alpha x}-2abe^{-\alpha x}+b^2 
\end{equation}
which can be recognized as the Morse type potential. It holds in the entire full-line. Employing the coordinate transformation  $y=\frac{1}{v_0 \alpha}e^{\alpha x}$ folds the interval $(-\infty, \infty)$ into the half-line $(0,\infty,)$. In new variable $y$ we thus obtain the Coulomb for the potential  
\begin{gather}
    V_\pm(y) = (l^2 \mp l)\frac{1}{y^2}-\frac{1}{y}+\frac{1}{4l^2}, \quad y \in (0,\infty,)
\end{gather}
where we have denoted $l=\frac{a}{v_0 \alpha}$ and $2b=\frac{1}{l}$.

To solve (5.9) we will be guided by the standard solution of the differential equation
\begin{gather}\label{geneqn}
    -\frac{d^2 \Lambda(s)}{ds^2}+\left(p^2+\frac{q}{s}+\frac{r}{s^2}\right)\Lambda(s)=0
\end{gather}
whose general solution is given by \cite{macd}
\begin{gather}\label{genSolCoulomb}
    \Lambda(s) \propto s^{\frac{1 \pm \sqrt{1+4r}}{2}} e^{-ps} {}_1U_1\left( \frac{q}{2p}+\frac{1 \pm \sqrt{1+4r}}{2}, 1 \pm \sqrt{1+4r};2ps \right)
\end{gather}
where ${}_1U_1$ is confluent hypergeometric function. Keeping in mind the convergence of the solution we take the positive root and assume the condition  
\begin{gather}\label{cond}
\frac{q}{2p}+\frac{1}{2}\Big(1+ \sqrt{1+4r}\Big)=-n \qquad 
\textit{where $n=0,1,2,......$ }
\end{gather}
to hold \cite{flug}. Comparing with ($\ref{geneqn}$) corresponding to the upper component, we obtain the relation
\begin{gather}
   p^2=\frac{1}{4l^2}-(E^2-m_0^2v_0^4), \quad q=-1, \quad r=l(l-1) \quad 
\end{gather}
In terms of the energy levels this translates to the result
\begin{gather}
   E^2_{+n}= m_0^2v_0^4+\frac{1}{4l^2}-\frac{1}{4(n+l)^2} \qquad 
\textit{where $n=0,1,2,......$ }
\end{gather}
for the upper component of the potential. These looks similar to what we have for the the usual hydrogen atom problem. The corresponding eigenfunctions follow from (\ref{genSolCoulomb}). Let us remark that the eigenfunctions can also be expressed in terms of the Whittaker function $M(k,m;z)$ \cite{mosh}   
\begin{gather}\label{shosol}
\Phi_{+}(y) \propto M\left(\frac{l}{\sqrt{1 - 4 \tau^2 l^2}}, l- \frac{1}{2} ; \frac{\sqrt{1 - 4 \tau^2 l^2} }{l}y\right)
\end{gather}
where $\tau^2= E^2-m_0^2v_0^4$.

To conclude this section, we give the result corresponding to the lower component of the potential in (5.11). We find for the energy values are
\begin{gather}
 E^2_{-n} = m_0^2v_0^4+\frac{1}{4l^2}-\frac{1}{4(n+l+1)^2}  \qquad 
\textit{where $n=0,1,2,......$ } 
\end{gather}
along with the associated eigenfunctions in terms of the confluent hypergeometric function
\begin{equation}
    \Phi_{-}(y) \propto y^{l+1} e^{-\frac{\sqrt{1-4\tau^2 l^2}}{2l}y} {}_1U_1\left( 1+l-\frac{l}{\sqrt{1 - 4 \tau^2 l^2}},  2l+2; \frac{\sqrt{1 - 4 \tau^2 l^2}}{l}y\right)    
\end{equation}
where $\tau^2=(E^2-m_0^2v_0^4)$. The presence of the exponential damping factor can be clearly noticed facilitating convergence in the region $-\frac{1}{2\tau} < l < \frac{1}{2\tau}$. We have thus solved the problem completely.

\section{Non-polynomial potentials}
From equation (\ref{pdmse}), (\ref{upperDirac}) and (\ref{lowerDirac}), we have for the two types of the wavefunctions the respective effective potentials
\begin{gather}
  V_{eff}^+= \Big(W^2+ v_f W' -\frac{1}{4}{v'_f}^2-\frac{1}{2}v_f v''_f\Big)  \\
V_{eff}^- =  \Big(W^2 - v_f W' -\frac{1}{4}{v'_f}^2-\frac{1}{2}v_f v''_f\Big)   
\end{gather}
Expressing $W(x)$ in terms of an auxiliary function $\zeta(x)$ i.e.  $W(x)= \zeta (x) - \frac{1}{2}v'_f$, the corresponding effective potentials take the forms
\begin{eqnarray}
 &&   V_{eff}^+= \zeta^2 - v'_f \zeta + v_f \zeta'-v_f v''_f \\
 &&   V_{eff}^- = \zeta^2 - \frac{d}{dx}(v_f \zeta)
\end{eqnarray}
These conform to the generalized forms of the partner effective potentials. It is interesting to note that the following combinations of the LFV  and  the  auxiliary  function namely respectively, $v_f= \frac{8}{2x^2+1}$ and $\zeta (x) = x$, provide for $V_{eff}^-$  potential
\begin{gather} \label{cprsPot}
    V_{eff}^- = x^2+8\frac{2x^2-1}{(2x^2+1)^2}
\end{gather}
holding for the upper component. $V_{eff}^+$ is exactly in the same as the CPRS potential \cite{cprs} proposed a few years ago in the context of.... In the constant mass case, the potential (6.5) supports the eigenvalues and wavefunctions 
\begin{gather}
    \xi_n =-3+2n,  \qquad n=0,3,4,5.... \\
    \Theta(x) = \frac{P_n(x)}{(2x^2+1)}e^{-\frac{x^2}{2}}
\end{gather}
where $P_n(x)$ are given in terms of the Hermite polynomials $H_n (x)$ by
\begin{gather}
    P_n(x)= 
\begin{cases}
    1, \qquad n = 0 \\
    H_n(x) + 4n H_{n-2}(x)+4n(n-3)H_{n-4}(x), \qquad \text{n=3,4,5...}           
\end{cases}
\end{gather}
In Figure 1 a sketch of $V^+_{eff}$ is shown.

On the other hand, for the upper component we have
\begin{gather} \label{doblewellPot}
    V_{eff}^+ =x^2+\frac{8}{(2x^2+1)}+\frac{32x^2}{(2x^2+1)^2}+\frac{256}{(2x^2+1)^3}-\frac{2048x^2}{(2x^2+1)^4}
\end{gather}
To the best of our knowledge, this version of an extended non-polynomial potential is new. It depicts a double-well potential. For the first two terms it conforms to the standard type whose exact solutions are given  by the confluent Heun functions. For a detailed account we refer to \cite{dong}. The additional terms act as damping effects whose significance is insignificant. The profile of (6.10) is sketched in  Figure 2. Some remarks on the qualitative analysis are in order. The local minimum points of this above potential is at $x_{min}=\pm 0.964633$ where the values of $V_{eff}^+(x_{min})=3.74132$ while the maixmum value $V_{eff}^+(x_{max})=264$ which is at $x_{max}=0$. 

\begin{figure}[h]
  \centering
  \begin{minipage}[b]{0.45\textwidth}
    \includegraphics[width=\textwidth]{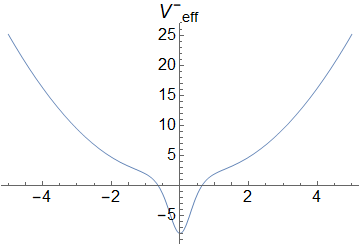}
    \caption{CPRS potential(\ref{cprsPot}).}
  \end{minipage}
  \hfill
  \begin{minipage}[b]{0.45\textwidth}
    \includegraphics[width=\textwidth]{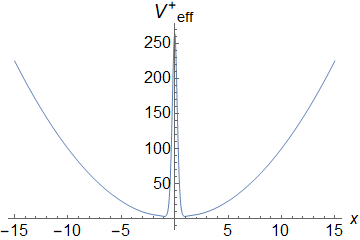}
    \caption{Double-well potential(\ref{doblewellPot}).}
  \end{minipage}
\end{figure}

\section{Summary}

To summarize, we have investigated in this article a class of Dirac Hamiltonian against the background of spatially-dependent mass and Fermi velocity. By converting to a pair of coupled equations when the Hamiltonian acts on the two-component spinor, we find that these equations resemble the corresponding PDM forms resulting from the von Roos prescription of the modified Hermitian kinetic energy operator. By observing that the underlying pseudoscalar potential also acting as the superpotential, we solve for several classes of solvable systems that include the free particle, shifted harmonic oscillator, 1D Coulomb and CPRS.

\section{Acknowledgment}

I thank Prof. Bijan Bagchi for valuable guidance.

\end{document}